\documentstyle[12pt,a4,epsfig,rotating,wrapfig]{article}

\begin{document}
\begin{center}
 
 
{\LARGE\bf The HERMES Back Drift Chambers} 

\vspace*{+1.5cm}

S.~Bernreuther$\,^a$, A.B.~Borissov$\,^b\,^1\,^2$,
H.~B\"ottcher$\,^b$, S.~Brons$\,^b$, \\ 
W.~Br\"uckner$\,^c$, A.~Buchsteiner$\,^b\,^3$, M.~Ferstl$\,^a$,
Y.~Gaerber$\,^b$, \\
A.~Gute$\,^a$, U.~Harder$\,^b$, D.~Hasch$\,^b$, M.~Kirsch$\,^a\,^4$, 
B.~Krause$\,^b\,^2$, \\ 
W.A.~Lachnit$\,^a$, F.~Mei$\ss$ner$\,^b$, G.~Modrak$\,^b$,  
F.~Neunreither$\,^a$, \\ 
W.--D.~Nowak$\,^b$, M.~Pohl$\,^b$, K.~Rith$\,^a$, H.O.~Roloff$\,^b$,
\\ H.~Russo$\,^a$, F.~Schmidt$\,^a$, A.E.~Schwind$\,^b$,
W.~Wander$\,^a\,^5$ \\  

\vspace*{+1.5cm}

\it{$\,^a$Universit\"at Erlangen--N\"urnberg, 91058 Erlangen, Germany
  \\ 
    $\,^b$DESY-IfH Zeuthen, 15735 Zeuthen, Germany  \\ 
    $\,^c$MPI f\"ur Kernphysik, 69029 Heidelberg, Germany \\
    $\,^1$on leave of absence from State University Saratow, Russia \\
    $\,^2$now at DESY, Deutsches Elektronen Synchrotron, 22603
    Hamburg, Germany \\
    $\,^5$now at Universit\"at Potsdam, 14415 Potsdam, Germany \\
    $\,^4$now at Firma "Dr. Bernd Struck", 22889 Tangstedt, Germany \\ 
    $\,^5$now at MIT, Cambridge, MA 02139, USA }

\vspace*{+4.0cm}


\begin{abstract}

The tracking system of the {\sc HERMES} spectrometer behind the
bending magnet consists of two pairs of large planar 6-plane 
drift chambers. The design and performance of these chambers is
described. This description comprises details on the mechanical and
electronical design, information about the gas mixture used and its
properties, results on alignment, calibration, resolution, and
efficiencies, and a discussion of the experience gained through the
first three years of operation. 

\end{abstract}

\end{center}
\newpage

 
\section{Introduction}

The {\sc HERMES} apparatus is based on a forward spectrometer with 
an internal gas target in the HERA lepton ring to investigate the 
spin structure of the nucleon by measuring doubly polarised 
lepton-nucleon scattering. A complete description of the {\sc HERMES}
spectrometer can be found in \cite{spectrometer}. The {\sc HERMES}
internal gas target is described in \cite{target}. 

The apparatus is divided into two halves installed above and
below the HERA beam pipes. Each spectrometer half is equipped with 
the same sequence of detectors. In front of the bending magnet
the tracking system is composed of two 3-plane microstrip gas counters 
just behind the target, a small 6-plane drift chamber that  was
commissioned in the third year of running, and a pair of
6-plane drift chambers near the magnet. Three proportional chambers
are mounted within the gap of the magnet.
They are used in the reconstruction of low momentum tracks
that do not pass the magnet completely. Behind the magnet, tracking is
based upon two pairs of large 6-plane drift chambers, the Back
Chambers (BC). 

\begin{figure*}[htb]
 \begin{center}
   \hspace*{-0.5cm}
   \epsfig{file=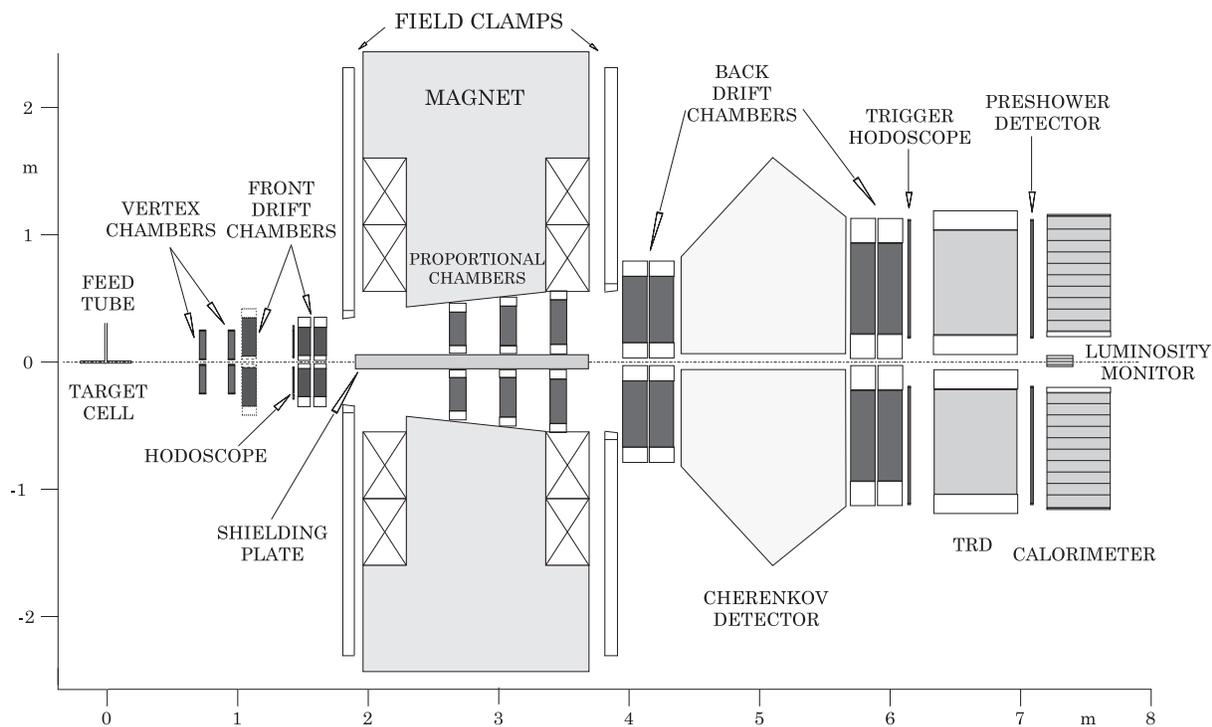, width=16.0cm, height=9.5cm, angle=0} 
   \caption{\label{hermes} \small Schematic diagram of the {\sc
       HERMES} spectrometer (side view).}
 \end{center}
\end{figure*}

The arrangement of the BCs within the {\sc HERMES} apparatus is
shown in Fig.~\ref{hermes}. Each track in the standard spectro\-meter
acceptance behind the magnet is measured by up to 24 individual planes
(up to 8 for each of the 3 wire orientations), providing high
redundancy. The most upstream pair of BCs contributes also to the
momentum determination for some of the short tracks, i.e. tracks that
leave the spectrometer prematurely behind the magnet due to their low
momentum. 

Particle identification is accomplished by a threshold Cherenkov
counter, a transition radiation detector, a preshower counter, 
and an electromagnetic calorimeter, all located behind the magnet. 
By combining the responses of these detectors, leptons and hadrons 
can be distinguished in the data analysis.

First results obtained from a test beam exposure of single BC modules 
have already been published elsewhere \cite{wcc95,Hasch}. Here 
further details on the construction of the Back Chambers and their 
performance in the experiment will be given.  

First data was taken in 1995 utilizing a polarised $^3$He target,
which to a good approximation can be considered as a polarised neutron 
target. In 1996 the target was changed to polarised H, which was used
throughout 1997 to investigate the spin structure of the proton. 
All results presented here are based on data taken in these first
three years of running.


\section{Mechanical Design}

The chamber modules consist of three pairs of wire planes with
alternating sense and potential wires between cathode foils. Both the
sense wire and cathode foil frames are made of GFK (glass
fiber-reinforced epoxy) by Stesalit (Zullwill, Switzerland) with a
thickness of 8~mm. Together with the sense wire spacing of 15~mm, this
leads to an almost square drift cell of size (15~mm x 16~mm). The
potential wires and cathode foils are at negative high voltage of
typically 1770~V, and the anode wires are at ground potential. The
wires are oriented vertically for the x--planes and at an angle of
$\pm$30$^{\circ}$ to the vertical for the u-- and v--planes. The
frames have sufficient length to allow both ends of all wires in the
u-- and v--planes to terminate on the long edges of the frames. Hence
all u-- and v--wires have the same length. To help solve the
left--right ambiguities, the planes of each pair are staggered by half
a drift cell. 

\begin{figure}[htb]
 \begin{center}
   \vspace*{-0.25cm}
   \epsfig{file=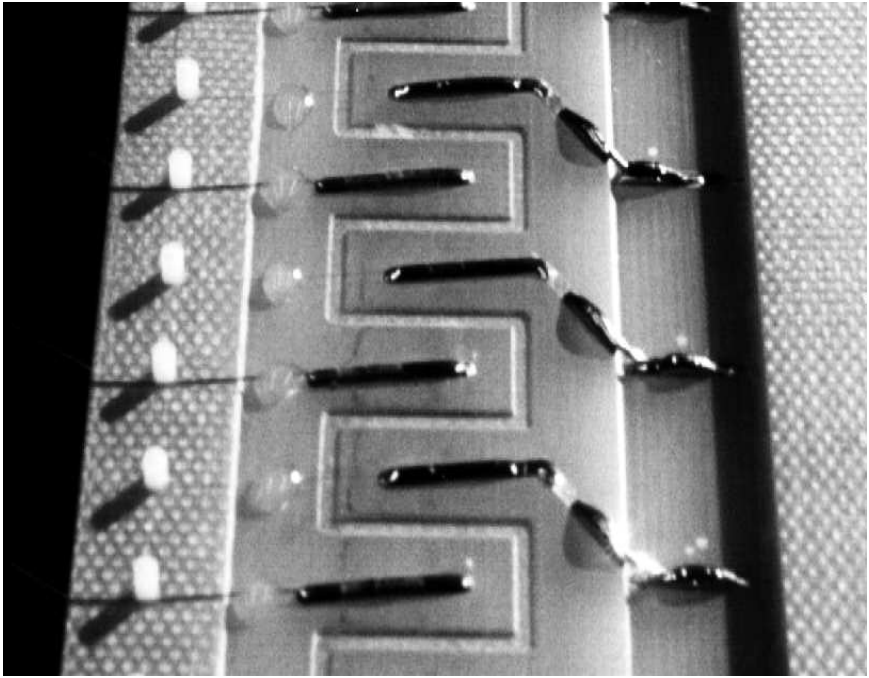, width=11.5cm, height=9.0cm, angle=0} 
   \caption{\label{wirepads} \small Picture of the soldering pads and
     positioning pins for the signal and potential wires. Seen also
     are the meander shaped groove between the pads and the connection
     of the signal wires to the readout line running beneath a GFK
     layer.} 
 \end{center}
\end{figure}

A high precision in wire placement is guaranteed by locating pins
made of polyoxymethylene. These pins are installed into the GFK-frame
along a curved locus that is deflected into a straight line by
prestressing before mounting the wires and foils. Therefore the
expected stress deflection of the frames caused by the wires and foils
is compensated in order to prevent displacement of the u-- and
v--wires. All wire positions were measured by an optical surveying 
system built for this purpose and were found to deviate less than
30~$\mu$m from the nominal positions \cite{Gute}. A special wire
running beneath a layer of GFK along one long side of each wire frame
and crossing under all signal traces provides capacitive
coupling of test pulses to check all channels of this plane. To avoid
surface currents between the soldering pads for signal and potential
wires, a meander shaped groove is machined between these pads. This
can be seen in Fig.~\ref{wirepads}, where the position pins and the
soldering pads are also visible. 

\begin{figure*}[htb]
 \begin{center}
   \vspace*{-1.75cm}
   \hspace*{-4.5cm}
   \psfig{file=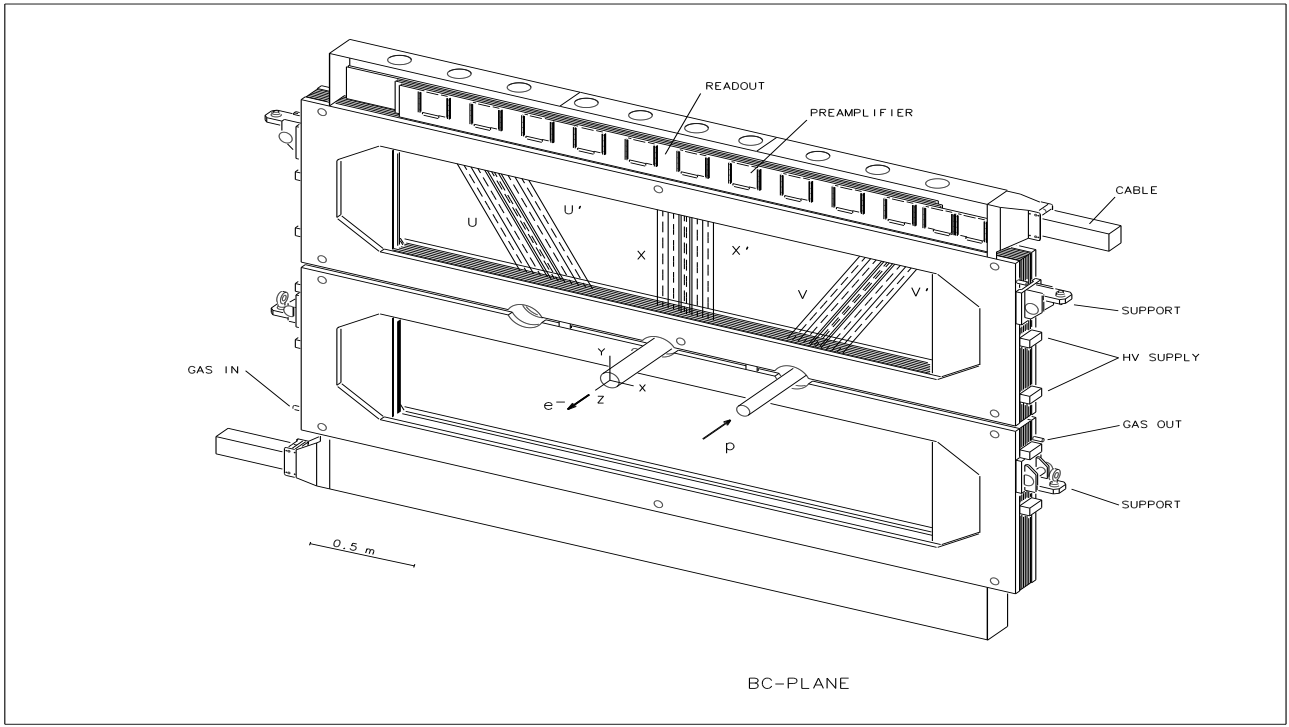,width=15.0cm,height=14.0cm,angle=-90} 
   \vspace*{-4.0cm}
   \caption{\label{bcmodule} \small Schematic view of the {\sc HERMES}
     BC module.}
 \end{center}
\end{figure*}

A schematic view of a complete BC module is shown in
Fig.~\ref{bcmodule}. In order to cover the {\sc HERMES} vertical
acceptance extending down to 40~mrad, the active areas of the drift
chambers had to be brought close to the beam pipe. This
is achieved by scallops in the frames for the beam pipes, as can be
seen in Fig.~\ref{bcmodule}. An additional requirement was the
rotational symmetry of the chambers around the lepton beam pipe, so
that each module can be mounted in the upper or in the lower half of
the spectrometer while retaining the same wire orientation. This is
the reason for the third scallop also seen in Fig.~\ref{bcmodule}. 
To guarantee the necessary stiffness and stability of the modules, the 
GFK--frames are mounted between two 38~mm thick frames made of
non--magnetic steel. The exact relative positioning of the wire planes
is defined by six steel bolts, each of 45~mm diameter, which run
through the stack of planes and are fixed in the steel frames. 

As the gas mixture used (see sect.~4) is heavier than air, the gas
stream into the chambers is directed from the bottom to the top
diagonally through the chambers (see Fig.~\ref{bcmodule}). The gas
seal is realized by O--rings made of neoprene--caoutchouc, between all
frames. To guarantee a smooth surface for this seal, the signal wire
traces pass under a GFK layer (see Fig.~\ref{wirepads}). The gas
leak rates are in the range expected from diffusion through the window
foils. 

The spectrometer layout requires two different sizes of BCs, the
smaller BC1/2 and the bigger BC3/4. The essential parameters of both
types are summarized in Table 1.

\begin{center}
Table 1. Back chamber properties  \\
\vspace*{+0.25cm}
\begin{tabular}{|l|c|c|} \hline
{\rule[-2mm]{0mm}{6.5mm}}Module active area (mm$^2$) & 
  1880 $\times$ 520 & 2890 $\times$ 710\\ 
{\rule[-2mm]{0mm}{6.5mm}}Sense wires per plane & 128 & 192\\ 
{\rule[-2mm]{0mm}{6.5mm}}Channels per module & 768 & 1152\\ \hline
{\rule[-2mm]{0mm}{6.5mm}}Cell width $\times$ gap & \multicolumn{2}{c|}
  {15 mm $\times$ 16 mm}  \\
{\rule[-2mm]{0mm}{6.5mm}}Anode wires &\multicolumn{2}{c|}  
  { 25.4 $\mu$m Au coated W}  \\
{\rule[-2mm]{0mm}{6.5mm}}Potential wires &\multicolumn{2}{c|}
  {127 $\mu$m Au coated Cu-Be}  \\
{\rule[-2mm]{0mm}{6.5mm}}Cathode foils & \multicolumn{2}{c|}
  {25 $\mu$m C coated Kapton}  \\  
{\rule[-2mm]{0mm}{6.5mm}}Rad. length per module &  \multicolumn{2}{c|} 
  {0.26 $\%$}    \\   \hline
\end{tabular}
\end{center}

\noindent
Further details can be found in 
Refs.~\cite{spectrometer,Bernreuther,Neunreither}. 


\section{Electronic Design}

The back chamber readout electronics is mounted directly on the long
edges of every module opposite to the beam pipes (see
Fig.~\ref{bcmodule}). Each electronics board contains 16
preamplifier-shaper-discriminator channels, consisting of a protecting
diode, a Fujitsu MB 43468 quad preamplifier, a common-emitter
post-amplifier, a fast comparator (MAX 9687), and an ECL gate used as
a driver. An external feedback at the comparator stage provides
hysteresis of 8~mV to reduce oscillation during the switching. The
overall transresistance is 60~mV/$\mu$A, given by the 20~mV/$\mu$A
transresistance gain of the Fujitsu cascade amplifier and an 
additional factor of three by the post-amplifier. 

To ensure electronic stability with the low thresholds required for
operation at low gas gain, all possible measures were taken to reduce
sensitivity to noise and cross-talk. LCL-filters ensure a clean low
voltage, and the threshold control voltage input is attentuated by a
factor of 100 by an operational amplifier circuit distributing it to
the comparators. The signal cable bundles are shielded to prevent
undesired feedback and all parts of the chamber ground system are
interconnected without leaving any gaps. Special materials are used to 
provide good electrical contact free of electrochemical corrosion.
These constructions form a high-continuity Faraday cage and
provide a common ground, especially for the front-end 
electronics and the high voltage input. This is important to prevent 
this system of 7680 channels from oscillating at low threshold values. 
At a voltage as high as 1800~V (above normal operating values),
the chambers exhibit a dark current less than the readout sensitivity
limit of 0.1 $\mu$A per plane. This behaviour is reflected in the fact
that there are almost no hot wires producing fake signals, and that
the dark rates agree with the expected cosmic ray background.

  \begin{wrapfigure}[17]{r}{8.0cm}
    \begin{center}
      \vspace*{-0.5cm}
      \epsfig{file=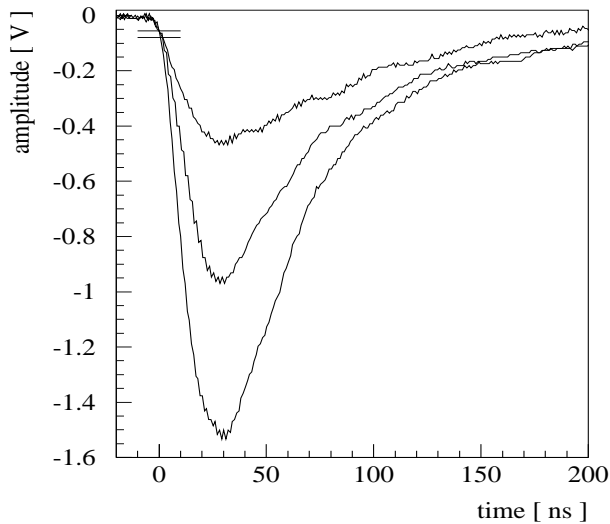, width=8cm, height=7cm, angle=0} 
      \caption{\label{pulse} \small Typical pulses and threshold
        values.} 
    \end{center}
  \end{wrapfigure}

Using rather low threshold values yields better resolutions as 
walk degrades the resolution for higher thresholds. This can be seen 
in Fig.~\ref{pulse} where three typical pulses are shown. 
The two horizontal lines indicate typically used threshold values.
Although even lower threshold values can be achieved,
they are not used, as crosstalk then would cause fake hits,
increasing the data volume and cost of event reconstruction. 
The crosstalk rate for low thresholds can be explained using the BC
raw signal amplitude distribution and the coupling between the
channels given by the quad preamplifier properties
\cite{Russo,Lachnit}. As a compromise, intermediate threshold values
were chosen, typically around 50~mV corresponding to an initial
current of 0.8~$\mu$A, or a charge of roughly $\sim~5.5 \cdot 10^4$
electrons. At these threshold values and high voltages of typically
around 1770~V used in the experiment, inter--plane crosstalk does not
play any role. 

Another possible impact on the quality of the data is crosstalk on
the 30~m long flat twisted pair cables running from the preamplifier
boards to the LeCroy 1877 {\sc Fastbus} TDCs located in the {\sc
  HERMES} electronics hut. The propagation times are perturbed by less
than 1~ns when all channels in a plane produce signals at the same
time. This is the situation for test pulse measurements but not for
physics events. Furthermore the routing of the ribbon cables is done
in a way providing maximal spatial distance between those channels
that usually fire simultaneously. Therefore, this effect is
negligible. 

The preamplifiers dissipate 400~W for the smaller and 600~W for the
bigger drift chamber modules, requiring efficient cooling. This is
provided by 9 big fans for each BC1/2 and 10 for each BC3/4 module. 
The temperature difference between the upper and lower parts of the
steel frame stays below two degrees. Different parts of the steel
frame exhibit slightly different thermal expansions due to this
temperature gradient. The gas temperature also differs between inlet
and outlet, which in principle could affect the
Space-Drift-Time-Relation (SDTR). However, neither of these thermal
effects degrade the chambers' resolution significantly.


\section{Gas Properties}

\begin{wrapfigure}[17]{r}{8.0cm}
    \begin{center}
      \vspace*{-1.5cm}
      \hspace*{+0.25cm}
      \epsfig{file=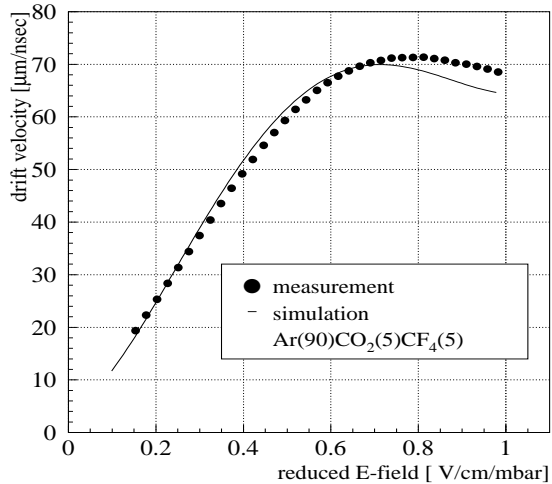, width=8cm, height=7cm, angle=0} 
      \caption{\label{vdrift} \small Drift velocity as function of the 
        reduced electrical field.}
    \end{center}
  \end{wrapfigure}

All drift chamber modules of the {\sc HERMES} spectrometer (see
Fig.~\ref{hermes}), including the BCs, are operated with the
same non-flammable gas mixture consisting of Ar(90~\%), CO$_2$(5~\%)
and CF$_4$(5~\%). 
Mixtures containing CF$_4$ have been shown to result in compromised
resolution due to electron attachment \cite{Biagi}. However, they 
have the advantages of short occupation time and long chamber
lifetime \cite{Openshaw}. The properties of this gas mixture were
measured with a small single-cell drift chamber designed to study
drift velocities and gas gain under varying conditions, including
changes in gas composition, gas contamination by N$_2$ and variations
in temperature and pressure \cite{Schmidt}. The measured drift
velocities $v_{\rm drift}$ vary between 20 and about 70~$\mu$m/nsec,
at reduced fields ranging between 0.15 and 1.0~V/cm/mbar. They agree
rather well with the results from simulations \cite{mag} as can be
seen in Fig.~\ref{vdrift}. 

Both CO$_2$ and CF$_4$ are used as quencher gases to absorb UV photons
while reducing the mean energy of the drifting electrons. The cross
sections for rotational and vibrational states of these gases are high
enough to 'cool' the electrons to energies close to the Ramsauer
minimum of the elastic cross section for argon, and the gas becomes
'fast' \cite{chris}. 
Especially CF$_4$ makes the gas fast and helps to avoid ageing effects
\cite{Openshaw}. After three years of running no indications of ageing
have been observed. 

The gas system supplying the drift chambers with the gas mixture was
built at TRIUMF/Canada. The system consists of a mixing station and a
recycler with an O$_2$/H$_2$O absorber. The composition of the gas,
the flow rates through each of the chambers and the recycled fraction
are programmable via a series of flow controllers. The gas volume of
each chamber is completely exchanged eight times per day. 
The composition of the outgoing gas is analysed by a Quadrupole Mass
Spectrometer. In addition, a small drift chamber is installed
in the outgoing gas stream to monitor continously the stability of gas 
gain and drift velocity \cite{Buchsteiner}.

  \begin{wrapfigure}[20]{r}{8.0cm}
    \begin{center}
      \vspace*{-1.5cm}
      \hspace*{+0.0cm}
      \epsfig{file=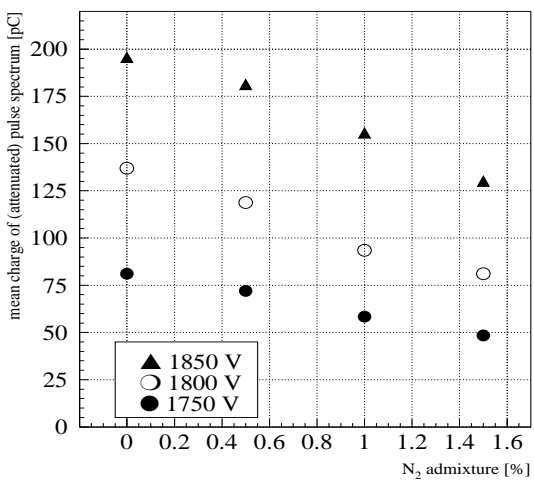, width=8cm, height=7cm, angle=0} 
      \caption{\label{n2_gain} \small Measured mean gas gain as
        function of the N$_2$ admixture at different high
        voltages. The vertical axis represents the mean values of the
        spectra from a charge--integrating ADC, and does not represent
        the absolute gas gain.} 
    \end{center}
  \end{wrapfigure}

Due to diffusion through the window foils of the BCs \cite{Lachnit}
there is contamination of the chamber gas by O$_2$, N$_2$ and 
H$_2$O, especially when operating the gas system in the recycling
mode. Whereas O$_2$ and H$_2$O are filtered out by the purifier, the
N$_2$ content depends on the recycled fraction.

The influence of the N$_2$ contamination on the drift velocity 
and the gas gain has been measured. An increasing content of N$_2$ 
results in only a slight decrease in the drift velocity. The gas gain,
however, is strongly reduced as can be seen from Fig.~\ref{n2_gain}
where the dependence of the gas gain on the N$_2$ admixture is shown
for three high voltages. Lower gas gain not only leads to a lower
chamber efficiency but also influences the chamber resolution, due to
increasing discriminator walk (cf. Fig~\ref{pulse}).
To prevent performance deterioration caused by an excessive N$_2$
contamination, the recycled fraction is set to 80~\% (20~\%
exhaust). This keeps the N$_2$ contamination stable at a level of
0.4~\% with negligible influence on both drift velocity and gas gain
(cf. Fig.~\ref{n2_gain}). 


\section{Alignment and Calibration}

A precise measurement of the particles' momentum requires careful
alignment and time calibration of all drift chambers involved in the
tracking system.

The BC geometrical alignment and time calibration is done in an
iterative procedure using reconstructed straight track segments
spanning the BC region. These track segments are part of a valid full
track. The tracks from all registered particles have been taken into
account, which are about 97~\% hadrons. From the
TDC information and the reconstructed hit position of the track in each
chamber plane, the residuals are calculated for the left
($r_{\mathrm{l}}$) and right ($r_{\mathrm{r}}$) half of the drift cell
separately. For a BC x--plane  this is given by the relation: 


\begin{equation}
r_{\mathrm{l(r)}} = +(-) \left[ x_{\mathrm{rec}}-x_{\mathrm{w}}
\right]-\frac{x_{\mathrm{d}}}{\cos\phi}. \label{resid} 
\end{equation} 

\noindent
where
$x_{\mathrm{w}}$ is the position of the wire fired,
$x_{\mathrm{rec}}$ is the position of the intersection of the
reconstructed track segment with the chamber plane, 
$x_{\mathrm{d}}$ is the drift distance calculated from the drift time
by using the SDTR, and $\phi$ is the angle of the track in the
xz-plane of the {\sc HERMES} spectrometer (for the {\sc HERMES}
coordinate system see Fig.~\ref{bcmodule}). The sign of the term
$\left[ x_{\mathrm{rec}} - x_{\mathrm{w}} \right]$ is {\it positive}
for the left and {\it negative} for the right half of the drift cell.  

From the mean value of these left/right residuals averaged over
many tracks, $\langle r_{\mathrm{l}} \rangle$ and $\langle
r_{\mathrm{r}} \rangle$, the alignment offset $\delta w$ and
the time offset correction $\delta t_{\mathrm{0}}$ of the trigger time
calibration are calculated according to

\begin{equation}
\delta w=\frac{1}{2}( \langle r_{\mathrm{l}} \rangle - \langle
r_{\mathrm{r}} \rangle ) \label{aligoff}
\end{equation}

\noindent
and

\begin{equation}
\delta t_{\mathrm{0}}=\frac{1}{2} \frac{ \langle r_{\mathrm{l}}
  \rangle + \langle r_{\mathrm{r}} \rangle}{\langle v_{\mathrm{drift}}
  \rangle} 
\label{t0off}, 
\end{equation}

\noindent
where $\langle v_{\mathrm{drift}} \rangle$ is the average drift
velocity.  

The calculation can be performed for a single drift cell, i.e. a
single wire, or for a whole plane by taking into account all drift
cells, i.e. all wires, of this plane. The variables relevant for a
whole plane will be denoted with $\overline{\delta w}$, the plane
alignment offset, and with $\overline{t_{\mathrm{0}}}$, the plane time
offset correction. 

%


In the following subsections the different corrections will be discussed
in more detail.

\subsection{Alignment}

During installation of the {\sc HERMES} spectrometer, the positions of
all detectors were measured by the DESY surveying group using optical
triangulation. The accuracy of this method is about 300~$\mu$m in the
transverse directions, i.e. in x and y, and of order 1.5~mm in
the longitudinal direction, i.e. in z. This is not sufficient for the
anticipated momentum resolution of the {\sc HERMES} spectrometer, which
requires the knowledge of the detector positions within at least
100~$\mu$m. The most important degrees of freedom in the alignment
procedure can be divided into two groups:

\begin{enumerate}
\item {\it Transverse alignment correction:} Simultaneous shift,
  perpendicular to the wire direction, of all wires in a given plane. 

  Because of the verified production accuracy of 30~$\mu$m for the
  uniformity of the wire spacing \cite{Gute}, this is the only 
  degree of freedom required for each plane in that coordinate which
  is in the wire plane, perpendicular to the wires. 

  This transverse alignment correction is calculated as the alignment
  offset $\overline{\delta w}$ from equation (\ref{aligoff}) by
  combining all wires in the given plane. The corrected position of
  the origin of this plane is used in the {\sc HERMES} geometry
  database for the next reconstruction cycle. After convergence of
  these iterations, the position of each plane is determined with an
  uncertainty smaller than 50~$\mu$m.  

  Some of these corrections of the individual plane positions can 
  result in an effective movement of the entire module. Especially for
  the u-- and v--planes, an alignment offset perpendicular to the wire
  orientation is related to a shift in both x and y. 
  Due to the degree of freedom in rotating the chambers around the
  beam axis, the entire chamber modules are shifted to minimize 
  deviations from the initial optical measurements.

\item {\it Longitudinal alignment correction:} Shift of a whole BC
  module in the z--direction. 

  For the correction of offsets in the z--position it
  is necessary to determine the dependence of the apparent x, u or v 
  alignment offsets on the incident track angle. Since alignment was 
  typically done with tracks recorded with magnetic field off, all
  tracks are straight and point back to the target interaction
  point. Hence the track angle is almost strictly correlated with the
  wire number in each plane, so that this angle dependence can not be
  seen in plane averages. Therefore the aligment offsets must be
  calculated for each individual wire or, at least, for several groups
  of wires. The track angle dependence arising from an offset in z is
  demonstrated in Fig.~\ref{zshift} for one specific plane of the
  lower BC module B3L. It is clear that, while using such a track
  sample, z alignment depends completely on being able to trust the
  absolute precision of the wire spacing. 

  \begin{figure}[htb]
      \begin{minipage}[l]{10.0cm}
      \vspace*{-1.25cm}
      \hspace*{+0.75cm}
        \epsfig{file=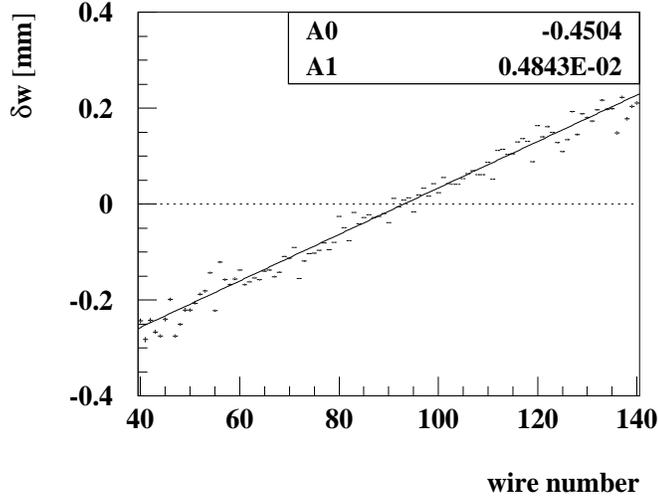,width=10.0cm,height=8.5cm,angle=0}  
      \end{minipage}
      \begin{minipage}[r]{4.0cm}
      \vspace*{-4.0cm}
      \hspace*{+1.25cm}
        \caption{\label{zshift} \small Transverse alignment offset
          $\delta w$ vs. wire number for a specific plane of the
          mo\-dule B3L located at z~=~5.795~m.}
      \end{minipage}
  \end{figure}

  The wire numbers are related to $x$ by the wire spacing. 
  Since all tracks come from essentially a point source at z=0, the
  slope of the fitted line can be represented by the relation 
  $\delta x / x = \delta z / z$. 
  For example, using this relation, a z-shift of 
  $\delta z$ = 1.87~mm is extracted for the plane illustrated in
  Fig.~\ref{zshift}. 

%

  \begin{figure}[htb]
      \begin{minipage}[l]{10.0cm}
       \vspace*{-1.0cm}
       \hspace*{+0.75cm}
       \epsfig{file=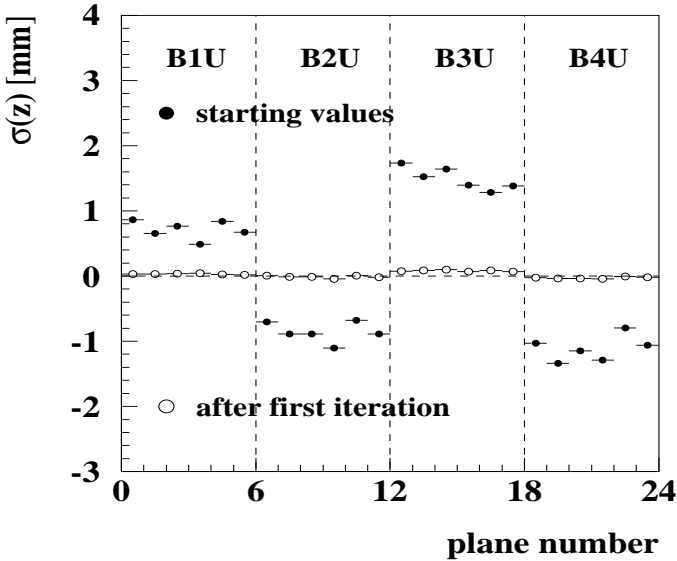, width=10.0cm,height=8.5cm,angle=0}
      \end{minipage}
      \begin{minipage}[r]{4.0cm} 
       \vspace*{-2.75cm}
       \hspace*{+2.25cm}
        \caption{\label{zplane}\small The calculated z--shift for all
          planes of the upper (B1U-B4U) BC modules. Shown are the
          starting values (full circles) and the result of the
          correction after one iteration step (open circles).} 
      \end{minipage}
  \end{figure}

  For each plane such a z--offset was calculated. The results for the
  upper (B1U -- B4U) BC modules are shown in Fig.~\ref{zplane} for
  the first two steps in the iteration procedure. 
  The results for the relative z--positions of the indivi\-dual planes 
  within a chamber module are consistent with design values and
  require no adjustment. Hence only a module z--offset was determined
  by averaging the individual plane z--offsets within a module, and
  the {\sc HERMES} geometry database was corrected accordingly for the 
  next iteration. This iterative procedure converges quite
  fast. Already after the first iteration the remaining z--offsets are
  below 200~$\mu$m (see Fig.~\ref{zplane}), and only one further step
  is necessary to reach the required accuracy.      

\end{enumerate}

Alignment corrections due to rotations of individual chamber modules
around certain axes have been investigated and are found to be
negligible \cite{Gaerber}. A relative top/bottom alignment is still
under investigation. 

\subsection{Time Calibration}

The signal times from the wires are measured by TDCs in common stop
mode. Correspondingly the earliest signal with the shortest drift
distance results in the largest TDC conversion value, and the actual
drift time is calculated as

\begin{equation}
t_{\mathrm{d}}=t_{\mathrm{0}}-t_{\mbox{\scriptsize{TDC}}},
\end{equation}

\noindent
where $t_{\mathrm{0}}$ is the trigger time offset depending on the
electronics configuration, the cabling, and the trigger
arrangement. It must be determined for each channel separately. For a 
$t_{\mathrm{0}}$ calibration based 
on track data, the data from several running periods (i.e. several
days) have to be combined to obtain sufficient precision. 
For those wires near the edge of the acceptance, even the combined
data sample did not deliver enough statistics and therefore 
the test pulse information had to be used to calibrate these wires
relative to the inner region of the same plane. 
Using equation (\ref{t0off}) the time offset correction
$\delta t_{\mathrm{0}}$ is calculated and the corrected
$t_{\mathrm{0}}$ value is loaded into the {\sc HERMES} calibration
database. To check the time stability of the $t_{\mathrm{0}}$
calibration, the plane averaged time offset correction
$\overline{\delta t_{\mathrm{0}}}$  is used. Time variations of the
$t_0$ values are caused either by changes in the hardware, e.g. in the
trigger system, or by changes in the running conditions of the
chambers, e.g. by variations of the atmospheric pressure.


\section{Track Residuals and Resolution}

The width of the residual distribution is a measure for the resolution
of a chamber plane. The residuals are calculated according to
Eq.(\ref{resid}). When quoting properties of the residual
distributions no distinction is made between the two halves of the
drift cell, i.e. the residual distributions for the left and right
half of the drift cell are combined. Residuals calculated for a single
drift cell will be denoted by $r$ and those averaged over the whole
plane by $\overline{r}$. 

To illustrate the relevance of the different alignment and calibration
steps discussed in the previous section, a system residual
distribution is produced by plotting the residuals calculated for the 
whole BC system, i.e. for all cells of all 

\begin{wrapfigure}[23]{r}{8.5cm}
    \begin{center}
      \vspace*{-1.75cm}
      \hspace*{+0.25cm}
      \epsfig{file=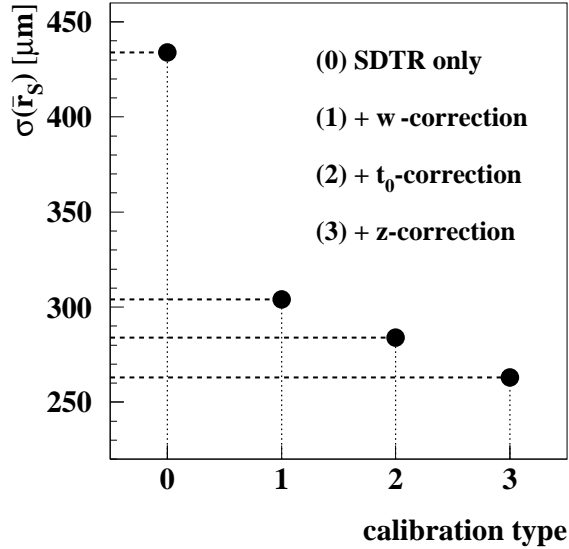,width=8.5cm,height=8.5cm,angle=0}
      \caption{\label{sysres} \small
        Width of the BC system residual distribution $\sigma
        (\overline{r_S})$ after the different alignment and
        calibration steps. The tracks from all registered particles
        are taken into account.}  
    \end{center}
\end{wrapfigure}

\noindent
planes of all modules. The width of this distribution, 
$\sigma (\overline{r_S})$, after each step is shown in
Fig.~\ref{sysres}. The improvement in width by more than 60~\%
demonstrates the necessity of the complete alignment and time
calibration procedure.  

\begin{wrapfigure}[23]{r}{8.5cm}
  \begin{center}
    \vspace*{-1.0cm}
    \hspace*{+0.25cm}
    \epsfig{file=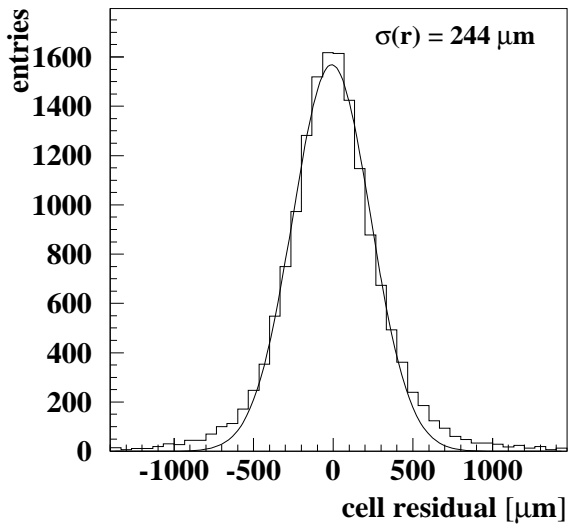, width=8.5cm,height=8.25cm,angle=0}
    \caption{\label{resol}\small A typical BC1/2 cell residual
      distribution calculated from the tracks of all registered
      particles (mainly hadrons). The curve represents a 
      Gaussian fit to the central region ($\pm$500~$\mu$m) of the
      distribution.}   
  \end{center}
\end{wrapfigure}

A residual distribution typical for a BC1/2 drift cell is shown in
Fig.~\ref{resol}. Aside from the tails, the distribution has the shape
of a Gaussian. Fitting its central region ($\pm$500~$\mu$m) with a
single Gaussian gives a mean value compatible with zero and a width
$\sigma(r)$ (see Fig.~\ref{resol}) typical for the BC1/2. 

\begin{wrapfigure}[20]{r}{8.25cm}
  \begin{center}
    \vspace*{-1.5cm}
    \hspace*{+0.25cm}
    \epsfig{file=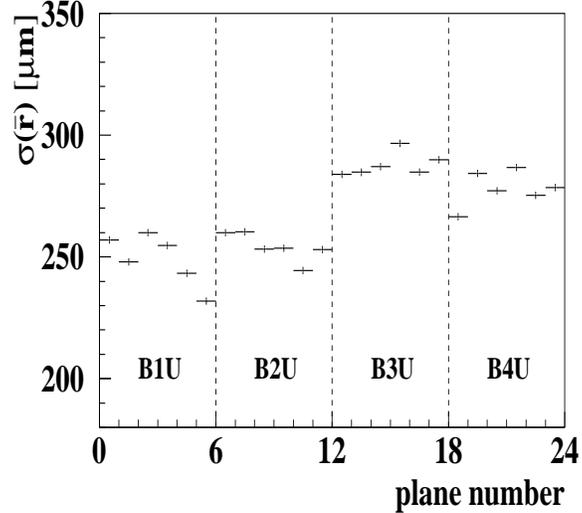,width=8.25cm,height=8.25cm,
      angle=0}  
    \caption{\label{bcresid}\small
      Width of the residual distribution $\sigma (\overline{r})$ of
      the upper (B1U-B4U) BC planes after all corrections as described
      in the text.}  
  \end{center}
\end{wrapfigure}

For each BC plane a residual distribution was calculated after having 
performed all the alignment and $t_{\mathrm{0}}$ offset
corrections. The resulting distributions were then fitted by a
Gaussian in the same manner as mentioned above. The widths of the 
residual distributions $\sigma (\overline{r})$ obtained for the
different planes of the upper BC modules under normal running
conditions are plotted in  Fig.~\ref{bcresid}. They are in the order
of 250~$\mu$m for the smaller BC1/2 modules and about 275~$\mu$m for
the bigger BC3/4 modules. Similar results were found for the lower BC
modules.  

\begin{wrapfigure}[22]{r}{8.25cm}
  \begin{center}
    \vspace*{-1.25cm}
    \hspace*{+0.5cm}
    \epsfig{file=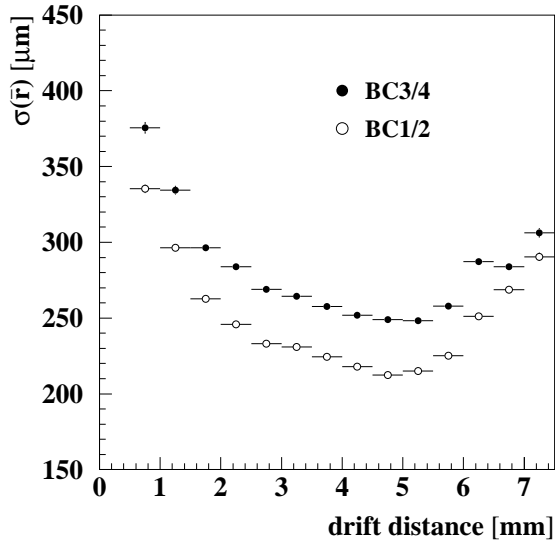, width=8.25cm,height=8.25cm,
      angle=0} 
    \caption{\label{cellresol}\small
      Width of the residual distribution $\sigma (\overline{r})$ of 
      the BC1/2 and BC3/4 as function of the drift distance. All
      corrections described in the text have been performed.} 
  \end{center}
\end{wrapfigure}

The width of the plane residual distribution for tracks crossing the
plane at an angle below 1$^{\circ}$ is given in Fig.~\ref{cellresol}
as function of the drift distance for both the BC1/2 and the BC3/4. 
The plot was obtained by taking into account the tracks of all
registered particles and fitting

\noindent
the peak region of the residual
distributions with a Gaussian. The tracks close to the sense wire
(i.e. small drift distances) were excluded because this region is
strongly affected by the difficulties in resolving the left/right
ambiguity in this calculation. However, a smooth behaviour of the
residual can be assumed. As expected, the best value is measured in
the central region of the drift cell to be about 210~$\mu$m for the
BC1/2 and about 250~$\mu$m for the BC3/4. The deterioration of the
spatial resolution for tracks near the sense wire (0~mm) is caused by 
the fluctuations of the primary ion-pair pro-

\noindent
duction statistics which leads in this region to
remarkable drift distance differences and for tracks near the
potential wire (7.5~mm) by electron diffusion \cite{Sauli}. In
addition, gas mixtures containing CF$_4$ have been shown \cite{Biagi} 
to affect the resolution due to electron attachment, which is
especially relevant for long drift distances.

Using the reconstructed track as reference track for calculating the
residual (`internal' method), the spatial resolution of a
chamber plane is related to the width of the residual distribution by
a geometrical factor that depends on the number of planes of each type
and their absolute position in z. This factor varies from plane to
plane by a few percent. For the {\sc HERMES} geometry the resolution
of a BC plane is about 10~\% bigger than the width of its residual
distribution \cite{Meissner}. 

For completeness we mention that smaller values for the resolution
had been obtained under test beam conditions and using an external
reference track defined by a Silicon Microstrip Detector telescope. A
value of 150~$\mu$m was measured in the central region of the BC drift 
cell \cite{wcc95,Hasch}. However, it should be noted that the $e^\pm$
momentum resolution in {\sc HERMES} is almost dominated by
bremsstrahlung in the materials of the target cell, the vacuum window,
and the vertex chambers (see Fig.~\ref{hermes}), and that the spatial
resolution of the BCs achieved under normal running conditions is
better than required for track reconstruction \cite{Wander}.  


\section{Plane Efficiency}

In the context of this paper the efficiency of a chamber plane is
defined as the fraction of reconstructed tracks for which a valid hit
from this plane was found in a certain corridor around this track. The 
corridor width adopted is the same as used in the reconstruction 
program \cite{Wander} to find all hits used to reconstruct the
track. It is about $\pm$900~$\mu$m for the back chamber planes,
corresponding to about $\pm$3~$\sigma$, where $\sigma$  is the plane
resolution discussed in the previous section. 

\begin{wrapfigure}[20]{r}{8.5cm}
 \begin{center}
   \vspace*{-1.25cm}
   \hspace*{-0.0cm}
   \epsfig{file=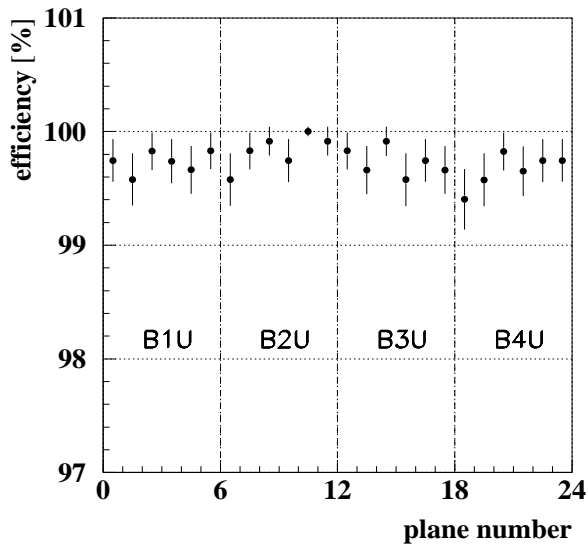, width=8.5cm, height=8.25cm,
     angle=0}
   \vspace*{-0.8cm}
   \caption{\label{planeeffi}\small 
     Plane efficiencies of the upper BC modules (B1U -- B4U)
     calculated for electron and positron tracks of one data run.} 
 \end{center}
\end{wrapfigure}

Obviously, the results from this definition depends on the
reconstruction algorithm. However, it has been checked that this
dependency does not introduce any significant bias. It is also obvious
that this measure does not account for other inefficiencies in track
reconstruction, which might arise from an excessive number of
extraneous hits, for example. 

As a typical example, the plane efficiencies of the upper BC modules
are shown in Fig.~\ref{planeeffi}, calculated for all identified
electron and positron tracks of one data run. 
It can be seen that for all planes this efficiency 
is high and the same within statistical errors. The overall average
plane efficiency for those tracks is determined to be well above
99~\%. For hadrons a lower efficiency is expected because of
their reduced ionization density in the chamber gas compared
to electrons and positrons. The efficiency is calculated to be about
97~\%  using the tracks of all registered particles, which are mainly
hadrons (see Fig.~\ref{effvsatm}).

The dependence of the plane efficiency on the drift distance, averaged
over all 

\begin{wrapfigure}[20]{r}{8.5cm}
 \begin{center}
   \vspace*{-1.25cm}
   \hspace*{-0.0cm}
   \epsfig{file=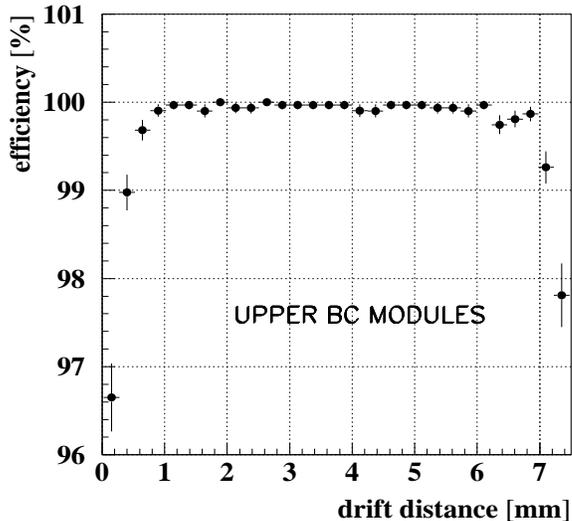, width=8.5cm, height=8.0cm,
     angle=0}
   \vspace*{-0.6cm}
   \caption{\label{celleffi}\small 
     Plane efficiency vs. drift distance for the upper BC modules,
     calculated for all registered tracks of one data run using a 
     wider corridor for calculating the efficiency than used for
     Fig.\ref{planeeffi}.}
 \end{center}
\end{wrapfigure}

\noindent
drift cells of all planes in the upper BC modules, is
displayed in Fig.~\ref{celleffi}.  
Here a wider corridor than the one mentioned above is used for
calculating the efficiencies, namely 1.5 drift cells, and
the tracks from all registered particles are taken into account.
In the central region of the drift cell the
efficiency is high and approaches almost unity, while towards the
sense wire (0~mm) and the potential wire (7.5~mm) it falls. This
behaviour is well known and can be explained by the same effects as
already mentioned in the previous section when discussing the drift
distance dependence of the resolution.

 
\section{Gain Stabilisation and Chamber Performance}

\begin{wrapfigure}[24]{r}{8.5cm}
 \begin{center}
   \vspace*{-1.25cm}
   \hspace*{+0.0cm}
   \epsfig{file=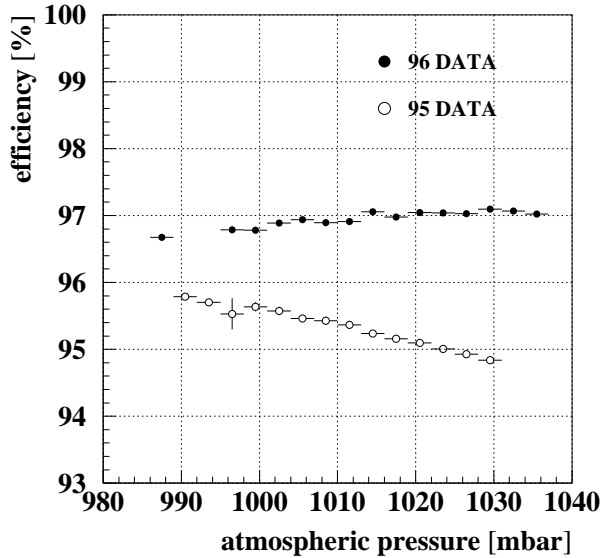, width=8.5cm, height=8.5cm,
     angle=0}
   \caption{\label{effvsatm}\small 
     Averaged BC plane efficiency as function of the atmospheric
     pressure for the 1995 data sample (open circles) and a 1996 data
     sample after introducing a gain stabilization scheme (closed
     circles). The efficiency is calculated from tracks of all
     particles, which are mainly hadrons.} 
 \end{center}
\end{wrapfigure}

On the basis of test run data and the early 1995 data analysis, an
optimal set of operating parameters (gas gain and electronic
threshold) was initially chosen for {\sc HERMES} running in order to
obtain high efficiency and good resolution at a moderate crosstalk. 
Studies of the time dependence of calibrations derived from the 1995
data indicate long term stability of the electronics. The expected
correlation between the chambers' performance and the atmospheric
pressure was also observed \cite{Lachnit}. As is well known, an
increase in pressure decreases the gas gain and vice versa. This was
the reason for introducing in the middle of 1996 running the dynamical
high voltage adjustment controlled by pressure measurements. The
'nominal' high voltage setting at normal atmospheric pressure of
1013~mbar was chosen to be 1770~V. In order to keep the gas gain
approximately constant during running, the high voltage is adjusted to
compensate for the change in atmospheric pessure using a
parametrisation given in \cite{Armitage}. The high voltage setting is
corrected in steps of 1~Volt over a range of $\pm$20~Volts, which
corresponds to a variation in pressure over a range of $\pm$30~mbar.  
This scheme improved the stability of the chamber performance
significantly, as illustrated by the reduction of the slope in
Fig.~\ref{effvsatm}. The improvement in average efficiencies by 1 to
2~\% achieved in 1996 compared to 1995, also seen in
Fig.~\ref{effvsatm}, is the result of further optimizing the wor-

\noindent
king conditions. In the {\sc HERMES} experimental environment,
threshold settings at the comparator input of 65~mV for BC1/2 and
80~mV for BC3/4 were found to ensure stable performance at low
crosstalk (cf. sect. 3). 

\begin{figure}[htb]
 \begin{center}
   \vspace*{-0.25cm}
   \epsfig{file=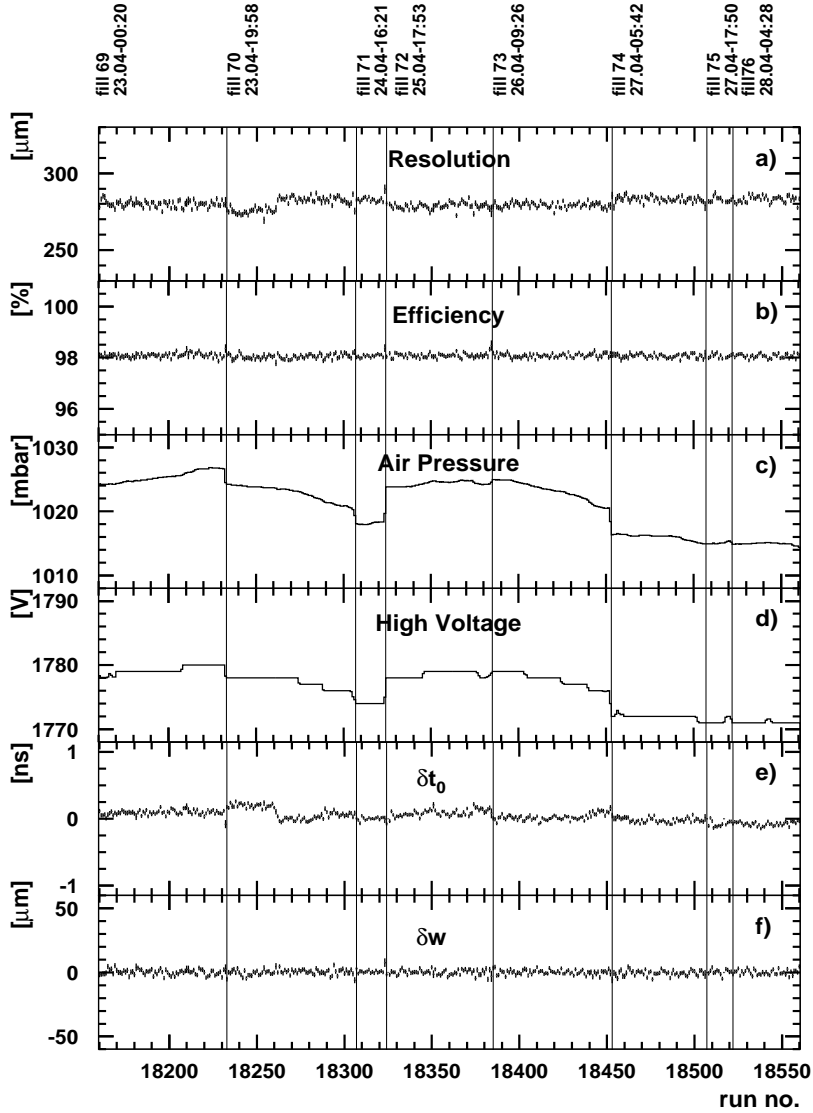, width=12.0cm, height=15.5cm,
     angle=0}
   \vspace*{-0.5cm}
   \caption{\label{performance}\small 
     Stability of the back chamber performance. Shown is the typical 
     behaviour of the resolution, the efficiency calculated from
     tracks of all registered particles (mainly hadrons), the
     atmospheric pressure, the high voltage, the residual time
     offset correction $\delta t_{\mathrm{0}}$, and the residual
     transverse alignment offset $\delta w$ for a specific BC plane
     over one week of 1997 running. The vertical lines indicate the
     fill boundaries.}    
 \end{center}
\end{figure}

The gain stabilization sche\-me, together with careful alignment 
and calibration work, resulted in remarkable long term stability of
the chamber performance. This can be judged from
Fig.\ref{performance}, which shows the behaviour of one specific BC
plane over one week of 1997 running. The vertical lines indicate the
beam fill boundaries. Typical fills of the HERA storage ring last
about 12 hours, separated by several hours for refilling. 

Resolution and efficiency were quite stable over this running period 
(see Fig.\ref{performance} a) and b)). The efficiency shown for
this particular BC plane turned out to be somewhat higher than the
average value for all planes shown in Fig.\ref{effvsatm}.
Variations in the atmospheric pressure were checked several times
per minute, and the high voltage was adjusted appropriately (see
Fig.\ref{performance} c) and d)). The stability of the 
electronics and alignment is shown in Fig.\ref{performance} e) and f). 
Plotted there are the remaining values deduced from the data for
$\delta t_{\mathrm{0}}$ (trigger time offset correction) and 
$\delta w$ (transverse alignment offset). The $t_{\mathrm{0}}$ time
offsets are calibrated once per fill. As can be seen, their variation
is below the level of the 0.5~ns time resolution of the LeCroy 1877
{\sc Fastbus} TDCs. The general alignment of the detectors is done
only once per year of operation. The residual alignment offsets
$\delta w$ can be used to monitor the stability of the alignment for
all runs. 

The back chambers have worked without major problems since the
commissioning of the experiment in spring 1995. They provided
very reliable tracking information for the back region of the {\sc
  HERMES} spectrometer.  
During the first three years of operation, no wire was broken and
only one wire lost connection by slipping out of the soldering. This,
however, did not degrade the tracking capabilities of the BC system in 
any significant way because of the high redundancy.  


\section{Online Monitoring}

To operate the BCs with high efficiency during routine operation, it
proved essential to have high quality online monitoring and slow
control capabilities. One of the event triggers (two tracks from
photoproduction) incorporates the x--planes of the most upstream upper
and lower BC modules and thus depends particularly on reliable and
stable operation. The low and high voltage supplies are integrated
into the {\sc HERMES} slow control scheme. All voltages are checked
several times per minute. Recovery from high voltage trips (typically
due to beam anomalies) is automatic. The temperature of the
electronics is checked by temperature sensors. All components of the
gas system are also accessible to the slow control system, and a
special monitor chamber is installed to check continuously the drift
velocity and the gas gain. A set of plots of distributions in hit
wires and drift times is provided for online checks.

An additional online drift chamber monitor program, 
based on a simplified track finding algorithm, provides performance
information for each chamber module independent of all other detectors
\cite{Meissner}. The combination of hits from the six planes of a
module is sufficient for the reconstruction of a useful sample of
tracks. A measure of the functionality of a module is the number of
tracks found using its data alone, compared to the average number of
tracks seen by all BC modules.
This relative rate is updated and displayed every few seconds to
monitor the functionality of the module.


\section{Conclusions}

The back chamber system proved to be one of the most stable and
reliable components of the {\sc HERMES} spectrometer. After careful
alignment and calibration work, the residual distributions of the BCs
have a width in the order of 250~$\mu$m ($\sigma$) for the smaller
BC1/2 and 275~$\mu$m for the bigger BC3/4 when calculated from the
tracks of all registered particles, which are mainly hadrons. The
chamber resolution is about 10~\% bigger than its residual width due
to a geometrical factor given by the {\sc HERMES} geometry. 
These numbers are considered to be acceptable in view of the chosen
gas mixture containing CF$_4$, which has shown to result in
compromised resolution due to electron attachment. 
The spatial resolution achieved is better than required for the
anticipated $e^{\pm}$ momentum resolution in {\sc HERMES}, which is
dominated by bremsstrahlung.
The single--plane detection efficiency for electron and positron
tracks is always above 99~\%.  
Gain stabilisation by dynamical high voltage control according to
atmospheric pressure leads to very stable operating characteristics
such as resolution, efficiency, and calibration.

Finally one can state that in its first three years of operation, the
back chamber tracking system has met all the requirements of the {\sc
  HERMES} experiment. 


\vspace*{1.0cm}

\noindent
{\bf Acknowledgements:} The authors are grateful to C.A.~Miller for
his continuous support of this project and for many stimulating
discussions. We very much appreciated the cooperation of R.~Openshaw
and M.~K\"uckes concerning the gas system. We are indebted to our
Russian colleagues A.~Laritchev, M.~Negodajev, W.~Kozlov, and
S.~Orfanitzky for their help during the assembly phase of the
chamber construction. Finally, we gratefully acknowledge the skillful 
and enthusiastic work of the mechanical and
electronics workshops in Erlangen and Zeuthen, which
was absolutely essential to build the whole system with high
quality and according to the schedule.



\end{document}